# Glide-Symmetric Acoustic Waveguides for Extreme Sensing and Isolation

Nikolina Janković[1] and Andrea Alù[2]

[1]*BioSense Institute-Research Institute for Information Technologies in Biosystems, University of Novi Sad, Dr Zorana Djindjica 1a, 21101, Novi Sad, Serbia*

[2]*Advanced Science Research Center, City University of New York, New York, New York 10031, USA*

*To whom correspondence should be addressed: nikolina@biosense.rs, aalu@gc.cuny.edu*

## Abstract

Glide symmetry offers new degrees of freedom to engineer the properties of periodic structures, and thus it has been exploited in various electromagnetic structures. However, so far there has been little exploration on the impact that glide symmetry can offer in the field of acoustics. In this paper, we explore glide-symmetric acoustic waveguides, highlighting their dispersion characteristics and guiding properties and demonstrating opportunities in the context of acoustic devices. Here we analytically derive their dispersive features applying a semi-analytical mode matching technique. We then demonstrate how the unusual dispersion properties of glide-symmetric acoustic waveguides can be used to achieve very sharp frequency responses. Based on these results, we propose a sensing platform for liquid analytes that exhibits large sensitivity and linearity. Furthermore, by introducing fluid motion, we leverage these responses to design an acoustic isolator based on acoustic Mach-Zehnder interferometry, whose design is more favorable in terms of footprint and complexity in comparison to other acoustic nonreciprocal devices that do not rely on glide symmetry.

## Keywords

Acoustic waveguide, glide symmetry, sensor, isolator, Mach-Zehnder interferometer

# Introduction

Periodic structures represent one of the most widely investigated and exploited classes of devices in electromagnetics [1-4], acoustics and mechanics [5-8] in the past decades, owing to their unique wave manipulation capabilities. Made of periodically arranged unit cells in 1-, 2- or even 3-dimensional space, they allow tailoring the dispersion characteristics and bandgaps, with useful functionalities to realize filters, slow-wave and non-linear structures, antennas, mirrors and sound focusing devices, and more recently topological structures [9] and non-reciprocal devices [10].

Although periodic structures have been investigated in various geometrical arrangements, they predominantly rely on discrete translational symmetries, which make them invariant under translation. Since such structures often exhibit strong dispersion and narrow bandwidths, in recent years structures with higher symmetries, including twist and glide ones, have been explored with the aim to tackle this intrinsic challenge [11]. Namely, a glide-symmetric structure coincides with itself after a translation and a reflection with respect to a so-called glide plane, thus providing an additional degree of freedom to engineer the properties of periodic structures. Glide symmetry was studied decades ago in terms of its application in waveguides [12-15], but only recently the interest in glide-symmetric structures has been revived. In [16-20] it was shown how such symmetry can be utilized to control and improve dispersion and bandwidth of standard guiding structures, including metallic waveguides and printed transmission lines. Moreover, glide symmetry has been used to improve gain and bandwidth of leaky-wave, slot array, helix and lens antennas [21-25], as well as the performance of phase shifters [26-28] and filters [29-30] realized in standard and groove gap waveguide technologies. It has also been shown that glide symmetry can be used to match the impedance of highly dense dielectric profiles over wide angles and broad bandwidths [31], and that glide symmetry can produce metasurfaces that exhibit high levels of anisotropy over wide frequency ranges [32].

Although glide symmetry shows a great promise to improve various device metrics, and although there is a strong analogy between electromagnetic and acoustic physics, glide-symmetric acoustic structures have not been explored so far, other than in [33] where a theoretical analysis of a meandering groove structure was analyzed. In this paper, we present a detailed analytical and numerical analysis of glide-symmetric acoustic waveguides in terms of their dispersion characteristics and guiding properties, and demonstrate their great potential for acoustic applications. To analytically derive the dispersive features of glide-symmetric waveguide we apply a mode matching technique [34-35], which is further supported by full-wave results obtained in numerical simulations. Furthermore, we use transmission line theory to provide an intuitive explanation of the forward-wave properties of glide-symmetric waveguides that have a negative slope in the band diagram [12, 17]. Finally, we demonstrate the advantages of the dispersion properties of the glide-symmetric waveguide by highlighting their potential for sensing and to implement a compact isolator based on an acoustic Mach-Zehnder interferometer combined with fluid motion.

**Theoretical background**

The geometry of interest and corresponding geometrical parameters are shown in Fig. 1. Without loss of generality, we consider a lossless two-dimensional (2D) geometry with perfect rigid surfaces, where the structure is invariant in the *y* direction and bounded in the *z* direction. The structure consists of two surfaces with corrugations of heights $h_1$ and $h_2$, which are periodic in the *x* direction with periodicity *d*, separated by a gap of height *g* and shifted by length *s* in the *x* direction. The three regions, lower and upper surfaces and the gap, are characterized by acoustic impedances and mass densities of the corresponding fluids given by $\eta_1$, $\rho_1$, $\eta_2$, $\rho_2$, and $\eta_{gap}$, $\rho_{gap}$, respectively. A glide-symmetric structure is obtained when $h_1 = h_2$, $\eta_1 = \eta_2$ and $s = \frac{d}{2}$, which corresponds to glide symmetry due to invariance under translation along *x* by half period and consequent mirroring over the *z* axis.

To derive the dispersion equation for the proposed structure in the general case, i.e., for $s \neq 0$, the mode-matching technique can be used, based on which the acoustic pressure and velocity fields in all three regions are expressed in terms of Floquet harmonics and matched using boundary conditions, providing a dispersion equation for this structure in matrix form. A detailed derivation of the dispersion equation can be found in Appendix A. To demonstrate the validity of our analytical formulation, Fig. 2(a) shows analytically and numerically obtained dispersion diagrams for glide-symmetric structures with $d = 100$ mm, $a = 80$ mm, $\frac{s}{d} = 0.5$, whereas the parameters $g$ and $h$ are varied. The fluid in all three regions is assumed to be air, with $\rho = 1.2 \frac{kg}{m^3}$, $\eta = 411.6 \frac{Ns}{m^3}$. Analytical results have been obtained from Eq. (A31) using Matlab for $p = 3$ and $m = 5$, while numerical results have been obtained using COMSOL Multiphysics. Excellent agreement between the results is found, which implies a high accuracy of the proposed analytical method for relatively small number of Floquet modes.

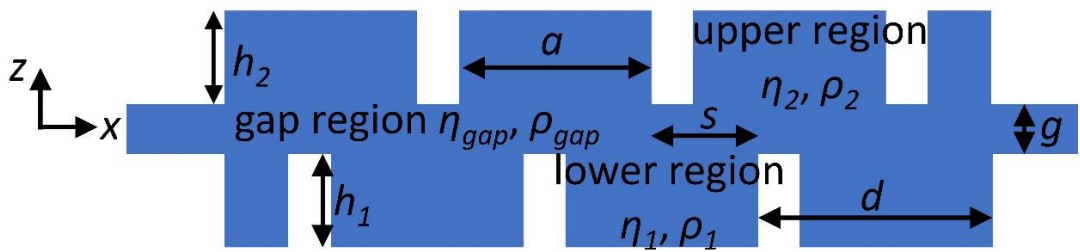

FIG. 1. Layout of the proposed structure.

The calculated dispersion diagrams also reveal the presence of two modes and no bandgap between them, which represents an interesting property of glide-symmetric structures - a broad frequency range with no stopbands in which propagation with small dispersion is enabled. Similarly to the electromagnetic scenario, this approach represents a promising avenue for controlling the propagation properties of standard acoustic waveguides. Nevertheless, this property is broken as soon as $\frac{s}{d} \neq 0.5$, and a bandgap opens between the two modes, as shown in Fig. 2(b).

Another interesting property of the proposed geometry can be noted in Figs. 2(a)-2(c): the geometrical parameters can be tailored to define the frequency ranges over which the structure allows propagation, but more notably they also have a strong effect on the dispersion itself. In particular, while the lower mode preserves its low-dispersive nature for different values of the geometrical parameters, the second mode can be tailored to exhibit larger dispersion and flattened quite considerably. Although this property may not be desirable for wideband guiding structures, we will demonstrate later how this property can be used to achieve compact structures for sensing and isolation applications. In addition, Fig. 2(d) shows the transmission coefficients for various geometries, accurately confirming the bandgaps arising in the corresponding dispersion diagrams.

It is interesting to observe that, as the branch corresponding to the guided mode in the waveguide is folded due to the introduced periodicity, the mode acquires a negative slope in the upper region of the dispersion diagrams, which would naively appear to indicate backward propagation. Interestingly, inspecting the field profiles in the waveguide, however, it is easy to confirm that the mode is actually forward propagating also in the upper branches, with a phase velocity parallel to the group velocity. Figs. S1 and S2 in [36] indeed show the propagation of the pressure acoustic field and acoustic analog of the Poynting vector for the lower and upper branches of the waveguide with $d = 100$ mm, $a = 80$ mm, $g = 20$ mm, $h = 30$ mm, and $s = \frac{d}{2}$, confirming that the behavior of the two modes is very similar, with a forward flow of phase as the wave propagates. This peculiarity has been noticed in electromagnetic glide symmetric structures, and was attributed to the coupling of space harmonics [12] or to the folding of the original unperturbed mode in the first Brillouin zone [17].

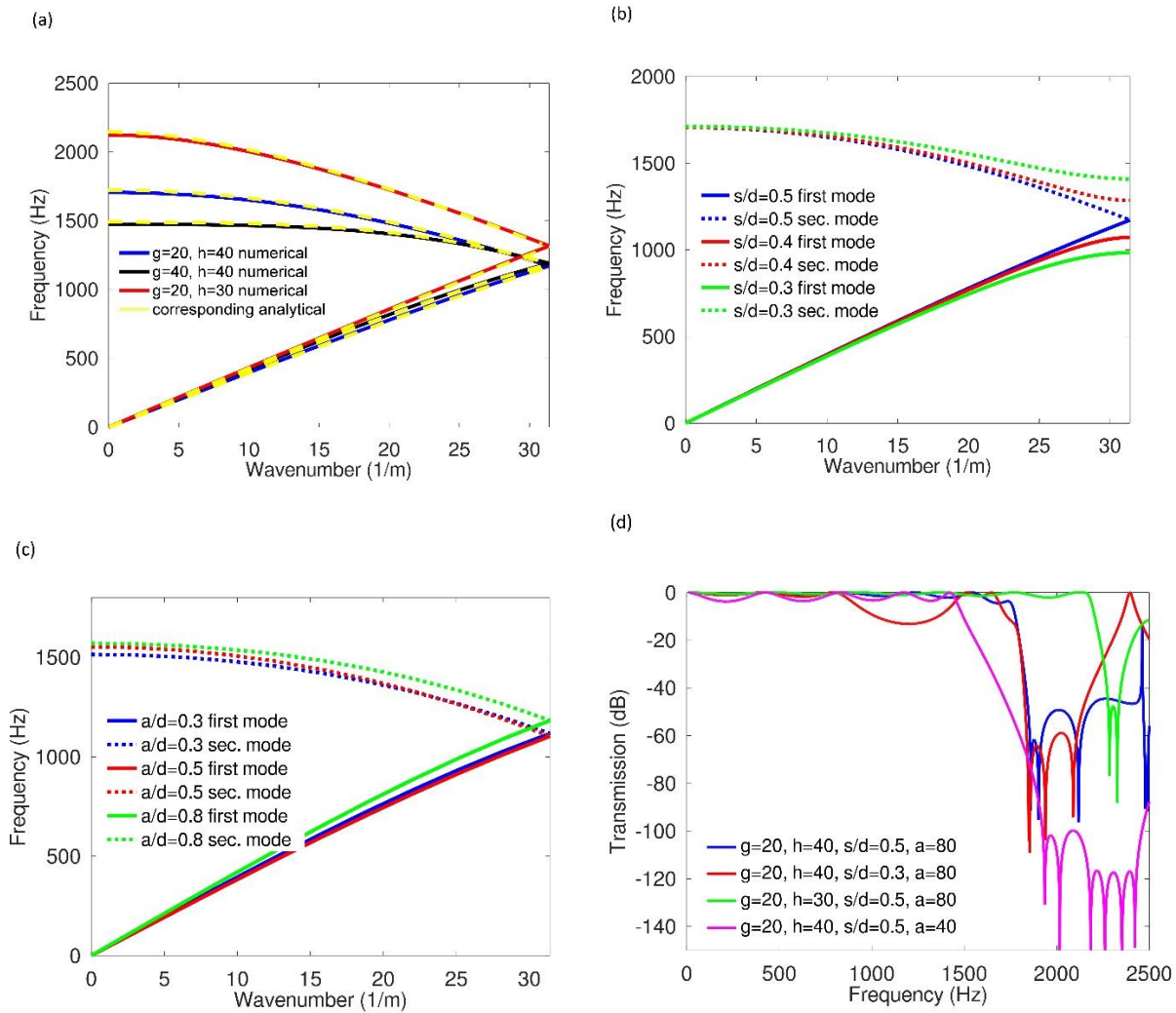

FIG. 2. (a) Dispersion diagrams for different values of parameters $h$ and $g$, $\frac{s}{d} = 0.5$, $a = 80$ mm. (b) Dispersion diagrams for different ratios $\frac{s}{d}$, $g = 20$ mm, $h = 40$ mm, $a = 80$ mm. (c) Dispersion diagrams for different ratios $\frac{a}{d}$, $g = 20$ mm, $h = 40$ mm, $\frac{s}{d} = 0.5$. (d) Transmission coefficients for different geometrical parameters. For all structures $d = 100$ mm. All values are given in mm.

In order to provide a more conclusive explanation for the negative slope of the upper branch of the dispersion diagram, we define the translation operator for period $d$ $T\psi(x,y,z) = \psi(x,y,z+d) = t\psi(x,y,z)$ where $T$ is the operator, $\psi$ is the eigenvector, and $t$ is the eigenvalue, whilst the definition of the glide operator is $G\psi(x,y,z) = \psi\left(-x,y,z+\frac{d}{2}\right) = g\psi(x,y,z)$, where $G$ is the operator, $\psi$ is the eigenvector, and $g$ is the eigenvalue [14]. After two consequent glide operations, one comes to the same result as one translation operator, i.e. $g^2 = t$, $g_{1,2} = \pm e^{-jk_x\frac{d}{2}}$, which implies that glide symmetry effectively reduces the periodicity of the structure by half [13],[37]. This implies that the first Brillouin zone becomes twice as large, indicating that the second mode in the dispersion diagrams in Fig. 2(a) is actually the mirrored replica over the axis $k_x = \frac{\pi}{d}$, and located in the region $\frac{\pi}{d} \leq k_x \leq \frac{2\pi}{d}$. That is why, the second branch actually is associated with a slightly perturbed forward mode.

This conclusion can be further supported using transmission-line theory and even and odd mode analysis [38]. To be able to correctly apply transmission line theory, we analyze a somewhat simplified structure with narrow corrugations, as shown in Fig. 3, which may illuminate on the origin of the negative slope. Due to symmetry, a unit cell of the simplified structure can be analyzed using even and odd equivalent circuits, as shown in Fig. 3. For the sake of simplicity, we consider that all parts of the structure have the same characteristic acoustic impedance $Z_0$, and thus the even and odd input impedances can be expressed as

$$Z_{ineven} = jZ_0 \frac{(\tan\theta_2 + \tan(t\theta))\tan((1-t)\theta) - 1}{\tan((1-t)\theta) + \tan\theta_2 + \tan(t\theta)}, \tag{1}$$

$$Z_{inodd} = jZ_0 \frac{\tan((1-t)\theta) + \tan(t\theta) - \tan((1-t)\theta)\tan\theta_2\tan(t\theta)}{1 - \tan(t\theta)(\tan((1-t)\theta) + \tan\theta_2)}, \tag{2}$$

where $\frac{s}{2} = t\frac{d}{2}$, $0 < t < 0.5$, $\theta = \beta\frac{d}{2}$, $\theta_1 = (1-t)\theta$, $\theta_2 = \beta h$, $\theta_3 = t\theta$, and $\beta$ is the wavenumber. The dispersion properties can be obtained using the expression

$$\beta = \frac{2}{d}imag\left(\tanh^{-1}\left(\sqrt{\frac{Z_{inodd}}{Z_{ineven}}}\right)\right). \tag{3}$$

Fig. 4(a) shows the analytically obtained dispersion diagrams for various values of the ratio $\frac{s}{d}$ and of the parameter $h$, while periodicity is the same for all cases, $d = 100$ mm. We find that in all geometries the first mode has a positive and the second mode a negative slope, which can be explained by the analysis of the key points marked as *A, B,* and *C* in Fig. 4. Namely, point *A* corresponds to the point at which $Z_{ineven}$ changes sign from negative to positive, *B* corresponds to the point at which $Z_{inodd}$ tends to infinity, while *C* corresponds to the point at which $Z_{inodd}$ changes sign from negative to positive. After some mathematical manipulation, these conditions can be formulated as

$$A: Z_{ineven} = 0 \Leftrightarrow \frac{2\cos\theta}{\sin((1-2t)\theta)+\sin\theta} = \tan\theta_2, \qquad (4)$$

$$B: Z_{inodd} \to \infty \Leftrightarrow \frac{2\cos\theta}{-\sin((1-2t)\theta)+\sin\theta} = \tan\theta_2, \qquad (5)$$

$$C: Z_{inodd} = 0 \Leftrightarrow \frac{2\sin(\theta)}{\cos((1-2t)\theta)-\cos(\theta)} = \tan\theta_2 \Leftrightarrow \frac{2\cos(\theta-\frac{\pi}{2})}{-\sin((1-2t)\theta-\frac{\pi}{2})+\sin(\theta-\frac{\pi}{2})} = \tan\theta_2. \qquad (6)$$

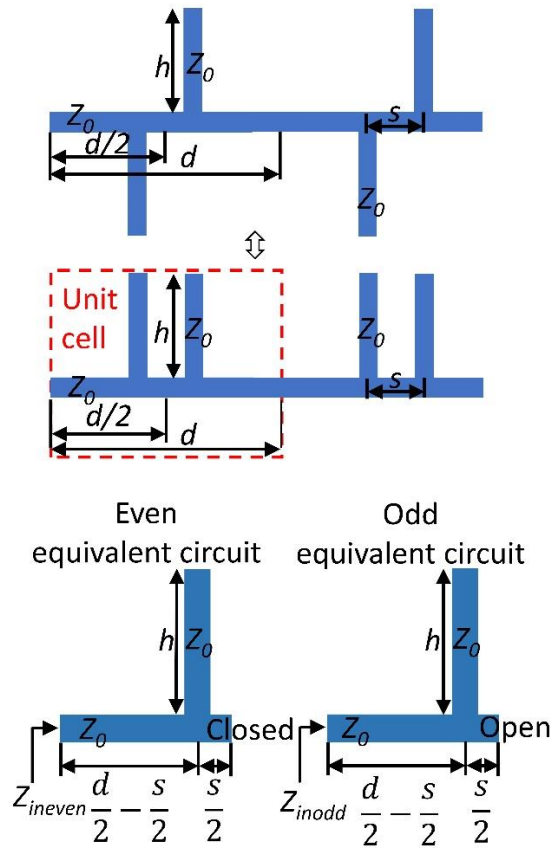

FIG. 3. Structure analyzed using the transmission line approach and its equivalent even and odd mode circuits.

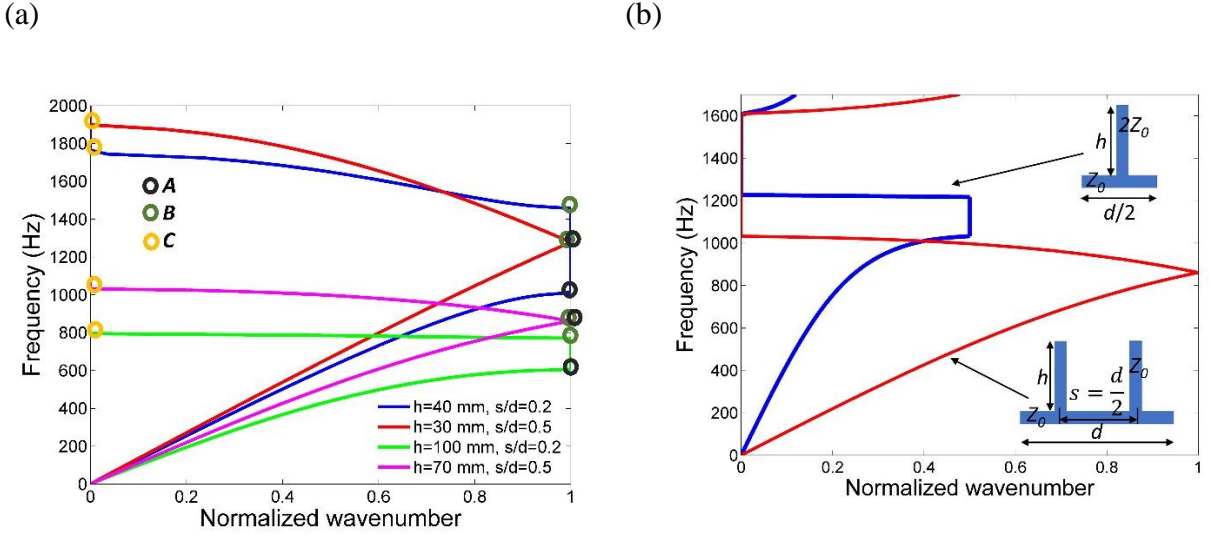

FIG. 4. (a) Dispersion diagrams for various scenarios of the ratio $\frac{s}{d}$ and parameter $h$ for the structure shown in Fig. 3. (b) Comparison of the dispersion diagrams for the structures with periodicity $d$ and $\frac{d}{2}$, $d = 100$ mm, $h = 70$ mm.

It can be seen that the parameter $t$ acts as an inverse coupling coefficient: the lower it is, the greater the distance between $A$ and $B$ in the spectrum, i.e., the broader the bandgap between the first and second mode. In the case $t = 0.5$, which corresponds to the glide-symmetric case, the conditions for $A$ and $B$ are identical, i.e. the points $A$ and $B$ are overlapped, and thus there is no bandgap between the first and second mode, which is a distinct feature of glide-symmetric structures.

Fig. 4(b) shows two dispersion diagrams for the structure of Fig. 3 in the glide-symmetric case, $s = \frac{d}{2}$. The first diagram corresponds to the case when the structure is treated as having periodicity $d$, the other assumes periodicity $\frac{d}{2}$, as indicated in the inset of Fig. 4(b). The two diagrams perfectly match each other in terms of the bandgap spectral locations, implying that the glide-symmetric structure has twice smaller periodicity and twice larger Brillouin zone. In other words, a physically-based justification as to why the branch slope is negative, yet the phase velocity of the underlying mode is positive can be provided considering the two geometries in Fig. 4(b), the glide symmetric waveguide and the corresponding periodic waveguide with period $\frac{d}{2}$. As mentioned above, the period of the structure is halved by the glide symmetry compared to the periodic scenario, implying that, even though the original (forward) mode supported by the periodic waveguide is only minorly perturbed by the additional loading every half period, the Bloch phase is now monitored every $d$. Therefore, for the portion of the forward mode with $\frac{\pi}{d} \leq \beta \leq \frac{2\pi}{d}$, the phase monitored every $d$ actually looks like advancing negatively, since over every period $d$ it accumulates more than a $\pi$ phase shift. Since the band diagram only defines phase velocities based on the Bloch phase at each period, we correctly find a negative slope for that portion of the branch. In reality, the microscopic distribution of the fields inside the unit cells reveal a different story, i.e., that the mode is simply a slow forward mode that accumulates in each period a long phase delay, rather than a fast backward mode. Therefore, our analysis confirms the interpretations in [13] and [37] in terms of the effective periodicity, but it also reconciles once and for all the forward nature of the upper branch in Fig. 2 with its negative slope in the dispersion diagram.

**Application as an acoustic sensor for liquid analytes**

Slow waves are well known for their excellent sensing potential, since their behavior is very sensitive to small changes of the parameters of the surrounding medium. Surface plasmon polaritons are one of the most prominent representatives of slow waves, and they have been widely explored in various sensing scenarios [39-42]. On the other hand, sensors for liquid analytes play an important role in environmental, agricultural, and medical applications due to the growing need for fast, reliable, and low-cost analysis of liquids such as water, saliva, blood etc. Whilst acoustic waves have been exploited in fluid manipulation and liquid sensing [43-48], glide-symmetric structures and their slow wave behavior have never been used for such purpose.

In order to demonstrate the advantages of the proposed structure, we present the use of our glide-symmetric device for sensing of liquid analytes, which employs its slow wave nature and the peculiar dispersion of the upper mode. Our sensing structure consists of three unit cells of dimensions $d = 100$ mm, $a = 80$ mm, $s = \frac{d}{2}$, $h = 35$ mm, and $g = 40$ mm and it overall length is equal to 530 mm, Fig. 5(a). The structure is filled with a mixture of water and glycerol. We have calculated the dispersion diagrams for several different mixtures, for which the liquid analyte was modelled using the parameters given in Table 1 [49], Fig. 5(b). The diagrams clearly indicate a slow wave response of the sensing structures and more importantly, they indicate an excellent sensitivity. Namely, if we consider the specific frequency of 6600 Hz, one can note that four dispersion diagrams significantly differ in wavenumbers, which implies that the signals travelling through the glide-symmetric structure with four different liquids would accumulate very different phase. The frequencies at which the bandgaps occur on the dispersion diagrams are also very distinguishable. Therefore, the sensing principle can be based on the accumulated phase shift as well as on the spectral shift of the transmission zero, which occurs due to the bandgap.

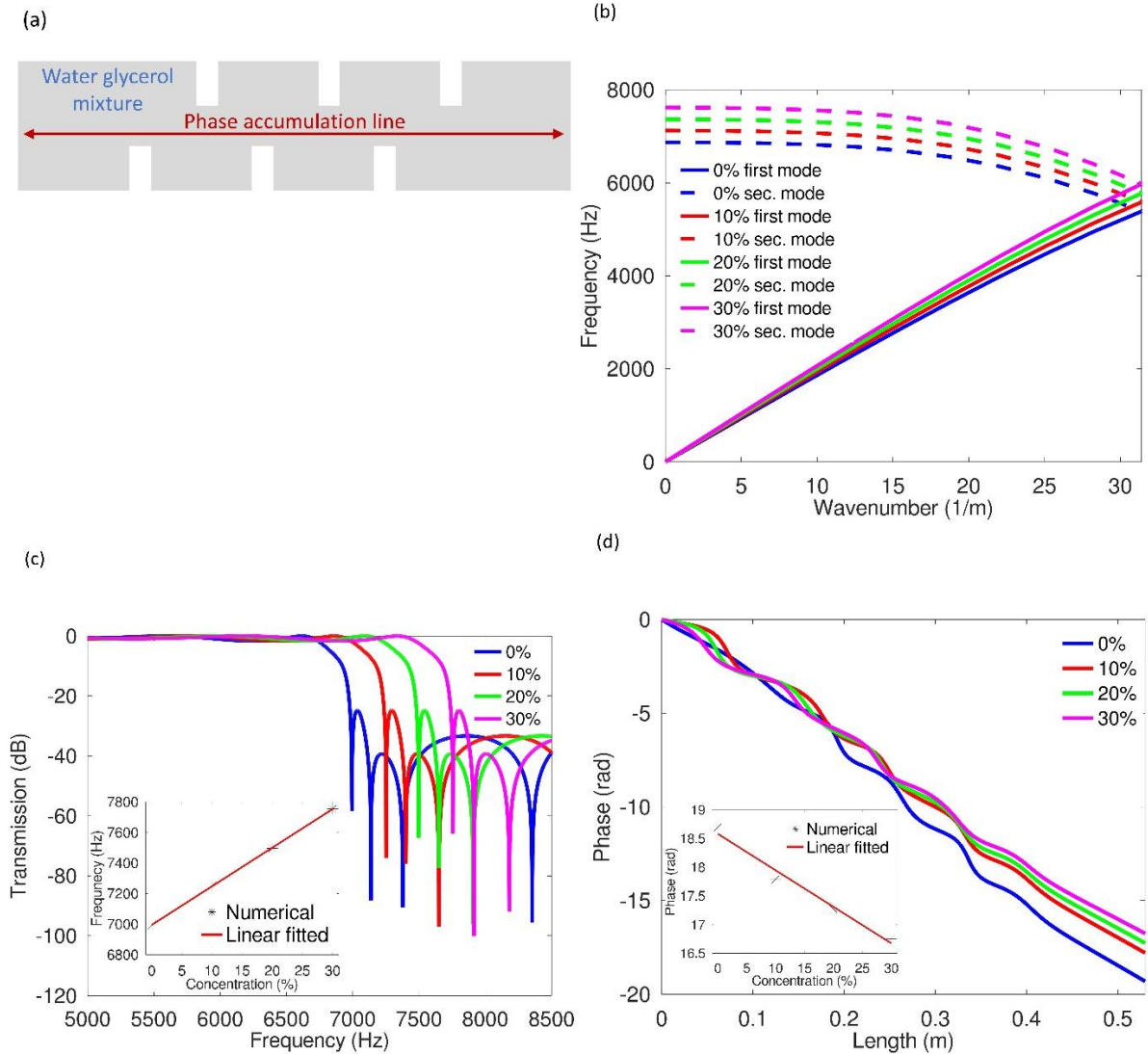

FIG. 5. (a) Sensor layout. (b) Dispersion diagrams, (c) Transmission responses, (d) Phase accumulation along the structure at 6600Hz, for different water glycerol mixtures.

The two sensing principles are illustrated in Figs. 5(c)-5(d), in which the responses have been obtained numerically for the same parameters as in the case of the dispersion diagrams. By monitoring the phase shift, phase accumulation was measured along the whole structure at 6600 Hz. In addition, the spectral position of the lowest transmission zeros in the response confirms the high sensitivity of the structure. The inset of Fig. 5(c) confirms a linear dependance of the transmission zero spectral position in terms of changes in the liquid parameters, while in Fig. 5(d) it can be seen that once glycerol is introduced in the mixture, the dependance becomes linear, i.e., the case of pure water somewhat deteriorates the linearity. The expected sensitivity is 25.37 Hz/%glycerol and 0.05 rad/%glycerol, in the case of the transmission zero and phase shift detection methods, respectively. Although these results are at the level of proof-of-concept, the presented sensor has a good potential to be further developed for various applications.

TABLE 1. Properties of glycerol-water mixtures (at 20 °C)

| Glycerol (wt.%) | Viscosity (cP) | Density (g/cm$^3$) | Ac. velocity (m/s) | Bulk modulus (GPa) |
|---|---|---|---|---|
| 0 | 0.91 | 1.002 | 1462 | 2.142 |
| 10 | 1.21 | 1.025 | 1516 | 2.355 |
| 20 | 1.62 | 1.046 | 1567.2 | 2.568 |
| 30 | 2.33 | 1.071 | 1621.2 | 2.814 |

## Application as acoustic isolator based on acoustic Mach-Zehnder interferometer

Another application that we demonstrate based on the same platform is a compact isolator incorporating the glide-symmetric waveguide, in which we also include fluid motion. Isolators are two-port devices that allow large transmission in one direction while suppressing the waves travelling in the opposite direction, and as such, their realization requires breaking reciprocity between a source and a receiver. Steady fluid flow has been shown to be an effective way to break reciprocity for sound [50-51] and recent work has shown how adding a slow material flow to a resonant device can make it highly nonreciprocal [52].

The underlying idea behind our approach to breaking reciprocity resides in fact that, in a perfectly rigid walled waveguide with transversely uniform flow, the phase velocity of a sound wave differs from the sound speed according to $c_{ph} = c \pm vk$ where $c$ is the speed of sound, $v$ the velocity of fluid flow, $k$ the wavevector direction, and the plus sign stands for waves travelling in the same direction as the flow, while the minus sign for the opposite case [53]. For low Mach numbers $\frac{v}{c}$, the wavenumbers of the modes are increased or decreased by $\frac{v\omega}{c^2}$ when sound propagates against or in the same direction as the flow. Since the wavenumbers of the two oppositely propagating sound waves differ, the two can be entirely cancelled by destructive interference after travelling through two waveguides of the same specific length and opposite fluid flows. Based on this concept, in [54] an isolator based on an acoustic Mach-Zehnder interferometer was proposed, and it was shown that a low Mach number flow through straight channels is sufficient to achieve a significant amount of isolation in waveguiding structures. Although such structure has significantly surpassed other acoustic isolators in terms of footprint area, the footprint can be further reduced by utilization of acoustic glide-symmetric waveguides, whose dispersion, unlike standard waveguides, can be tailored to achieve very slow speeds.

In this sense, Fig. 6(a) shows dispersion diagrams for two acoustic glide-symmetric waveguides, one in which a steady fluid flow is applied and the other one with no fluid flow. Air is used as the fluid, while the fluid flow velocity is equal to 3.2 m/s to ensure low Mach number, the same as in [54]. The geometrical parameters of the waveguides are $d = 100$ mm, $a = 60$ mm, $s = \frac{d}{2}$, $h = 35$ mm, and $g = 40$ mm and they have been chosen to achieve isolation at approximately the same frequency as the structure proposed in [54].

The structures have been optimized to achieve larger wavenumber difference than in the case of standard waveguides for the same fluid flow velocity. As it was previously stated, the glide-symmetric structure acts as a periodic structure that exhibits slow wave properties, which implies a more pronounced slope of the dispersion diagram in the frequency region close to the bandgap. Therefore, the wavenumbers of the two glide-symmetric structures are significantly different at a specific frequency, than the two counterparts of a standard waveguide, Fig. 6(a). Fig. 6(a) also shows the dispersion diagram for a glide symmetric waveguide without fluid flow, whose dimensions are all the same, except for the gap which is now equal to 35.8 mm (green line). This dispersion diagram crosses the one of the structure with fluid flow at the desired operation frequency of 1550 Hz, i.e., the two wavenumbers are equal at the given frequency.

The layout of the proposed isolator, Fig. 6(b), reveals how these dispersion properties of these waveguides can be utilized to realize a compact acoustic isolator based on Mach-Zehnder interferometry: the input signal is divided into two branches – the upper one with a glide-symmetric structure with $g = 40$ mm and steady fluid flow of velocity 3.2 m/s in the positive $x$ direction, and the lower one with a glide-symmetric structure with $g = 35.8$ mm and no steady fluid flow. After propagation through the two branches, the signals are combined at the output port. If the signal propagates in the positive *x* direction, there is no phase difference between the two branches at 1550 Hz, since the corresponding wavenumbers are the same as indicated in Fig. 6(a). On the other hand, for a signal propagating against the fluid flow, there is a phase difference between the two branches, which is proportional to $2L(k_{fluidflow} - k_{nofluidflow})$, where *L* stands for the physical length of the branches, and $k_{fluidflow}$ and $k_{nofluidflow}$ correspond to the wavenumbers of the structures with *g* = 40 mm, with and without fluid flow. Interestingly, despite the fact that the wave number is small just looking at the band diagram, implying a very fast wave in terms of Bloch band, the group velocity is actually very slow, offering unique opportunities for large nonreciprocal responses with modest flow bias.

The length of the branches has been optimized to achieve a phase difference equal to $\pi$ for the signals propagating against the fluid flow, i.e., to achieve cancellation of the two signals at the output port at 1550 Hz. The final device features 12 unit cells in each branch, whereas their length *L* is equal to 1390 mm. The overall device area occupies 2390 mm x 400 mm, which is equal to $10.8\lambda_0$ x $1.81\lambda_0$, where $\lambda_0$ represents wavelength in air.

The response of the final structure was calculated using COMSOL Multiphysics. Due to optimization, the final operating frequency is 1552 Hz, and at this frequency high transmission for forward wave is obtained as well as excellent isolation, expressed as $20\log\left|\frac{S_{21}}{S_{12}}\right|$, of over 55 dB, Fig. 6(c). This is further confirmed by the acoustic pressure field distribution at the output ports for forward and backward signals, which indicates no phase difference and phase difference equal to $\pi$, respectively, Fig. 6(d). In comparison to the isolator presented in [54], the advantage of the proposed structure is twofold: the overall footprint is significantly smaller, $10.8\lambda_0$ x $1.81\lambda_0$ compared to $18\lambda_0$ x $4\lambda_0$ at the same operating frequency and for the same fluid flow velocity, and there is no need for reciprocal phase shifters, which makes the proposed isolator more favorable in terms of dimensions and complexity.

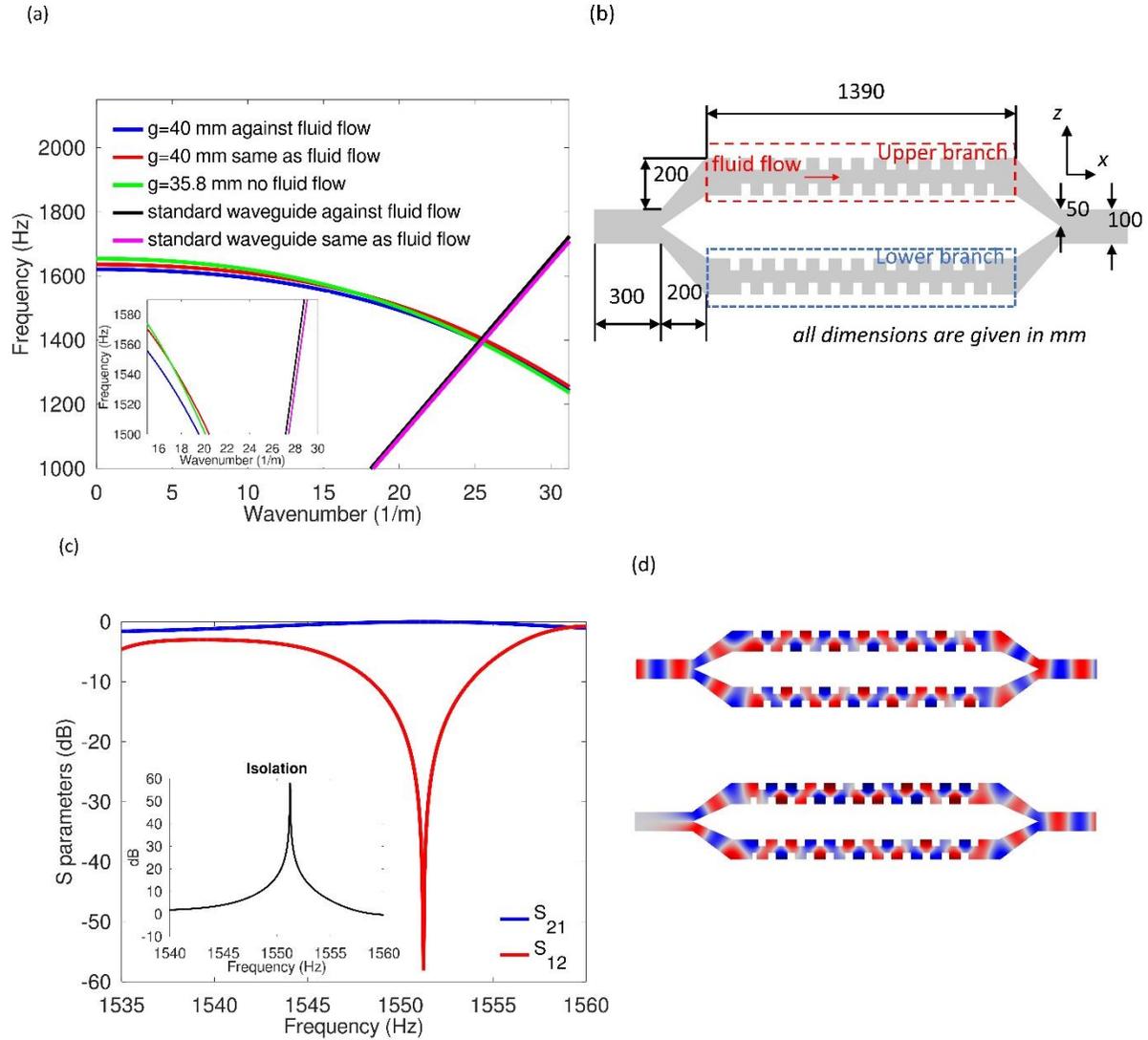

FIG. 6. (a) Dispersion diagrams for the structures with and without fluid flow. (b) Layout of isolator based on acoustic Mach Zehnder interferometer; all dimensions are given in mm. (c) Response of the isolator. (d) Acoustic pressure field distribution for forward and backward waves.

## Conclusions

In this work, we have investigated the response of an acoustic waveguide with glide symmetry and suggested its applications as a sensor and a compact nonreciprocal device. Using mode-matching and numerical simulations, we have demonstrated that glide symmetry provides a new degree of freedom to tailor the dispersion and guiding properties. After providing physical insights into the nature of the folded branch, in particular regarding its negative dispersion and slow-wave properties, we have discussed its potential to implement acoustic sensors and an acoustic isolator based on Mach-Zehnder interferometry. We have shown large sensitivity and lower complexity and footprint in comparison to other acoustic devices operating under similar principles but without the benefits stemming from glide symmetry.

## Acknowledgements

The work described in this paper is conducted within the project NOCTURNO, which receives funding from the European Union's Horizon 2020 research and innovation programme under Grant No. 777714. This work is also partially supported by the Simons Foundation and the National Science Foundation.

## Appendix A. Derivation of the dispersion equation for the acoustic glide-symmetric waveguide

Considering propagation in the positive *x* direction, and following the general expression for Floquet modes [34-35], the velocity and pressure fields in the gap region can be expressed as follows

$$v_{gapz} = \frac{1}{d}\sum_{p=-\infty}^{\infty}\left(A_p \sin(k_{z,p}z) + B_p \cos(k_{z,p}z)\right) e^{-jk_{x,p}x}, \quad (A1)$$

$$v_{gapx} = \frac{1}{d}\sum_{p=-\infty}^{\infty}\frac{-jk_{x,p}}{k_{z,p}}\left(B_p \sin(k_{z,p}z) - A_p \cos(k_{z,p}z)\right) e^{-jk_{x,p}x}, \quad (A2)$$

$$p_{gap} = \frac{1}{d}\sum_{p=-\infty}^{\infty}\frac{-j\eta_{gap}k_{gap}}{k_{z,p}}\left(B_p \sin(k_{z,p}z) - A_p \cos(k_{z,p}z)\right) e^{-jk_{x,p}x}, \quad (A3)$$

where the fields are related through the expressions

$$\nabla p_{gap} = -j\omega\rho_{gap}v_{gap}, \quad (A4)$$

$$\nabla \cdot v_{gap} = -j\frac{k_{gap}}{\eta_{gap}}p_{gap}. \quad (A5)$$

Angular frequency is denoted by $\omega$, while $p$ is an integer number, $A_p$ and $B_p$ unknown amplitude coefficients, $k_{gap}$ fluid wave number, $k_{x,p} = k_{x,0} + \frac{2\pi p}{d}$ and $k_{z,p} = \sqrt{k_{gap}^2 - k_{x,p}^2}$. We note here that $z = 0$ in the middle plane of the gap region.

Since the lower and upper corrugations can be regarded as acoustic waveguides of the length $h_1$ and $h_2$, respectively, which are closed at $z = h_1 + \frac{g}{2}$ and $z = -h_2 - \frac{g}{2}$, the corresponding fields are

$$v_{1z} = \sum_{m=0}^{\infty} C_{1m} \sqrt{\frac{2-\delta_{m0}}{a}} \cos\left(\frac{m\pi}{a}x\right) \left(e^{+jk_{1z,m}(z+\frac{g}{2})} - e^{-j2h_1 k_{1z,m}} e^{-jk_{1z,m}(z+\frac{g}{2})}\right), \quad (A6)$$

$$v_{1x} = \sum_{m=0}^{\infty} \frac{m\pi}{a} \frac{1}{-jk_{1z,m}} C_{1m} \sqrt{\frac{2-\delta_{m0}}{a}} \sin\left(\frac{m\pi}{a}x\right) \left(e^{+jk_{1z,m}(z+\frac{g}{2})} + e^{-j2h_1 k_{1z,m}} e^{-jk_{1z,m}(z+\frac{g}{2})}\right), \quad (A7)$$

$$p_1 = \sum_{m=0}^{\infty} -\frac{\eta_1 k_1}{k_{1z,m}} C_{1m} \sqrt{\frac{2-\delta_{m0}}{a}} \cos\left(\frac{m\pi}{a}x\right) \left(e^{+jk_{1z,m}(z+\frac{g}{2})} + e^{-j2h_1 k_{1z,m}} e^{-jk_{1z,m}(z+\frac{g}{2})}\right), \quad (A8)$$

$$v_{2z} = \sum_{m=0}^{\infty} C_{2m} \sqrt{\frac{2-\delta_{m0}}{a}} \cos\left(\frac{m\pi}{a}(x-s)\right) e^{-jk_{x,0}s} \left(e^{-jk_{2z,m}(z-\frac{g}{2})} - e^{-j2h_2 k_{2z,m}} e^{+jk_{2z,m}(z-\frac{g}{2})}\right), \quad (A9)$$

$$v_{2x} = \sum_{m=0}^{\infty} \frac{m\pi}{a} \frac{1}{jk_{2z,m}} C_{2m} \sqrt{\frac{2-\delta_{m0}}{a}} \sin\left(\frac{m\pi}{a}(x-s)\right) e^{-jk_{x,0}s} \left(e^{-jk_{2z,m}(z-\frac{g}{2})} + e^{-j2h_2 k_{2z,m}} e^{+jk_{2z,m}(z-\frac{g}{2})}\right), \quad (A10)$$

$$p_2 = \sum_{m=0}^{\infty} \frac{\eta_2 k_2}{k_{2z,m}} C_{2m} \sqrt{\frac{2-\delta_{m0}}{a}} \cos\left(\frac{m\pi}{a}(x-s)\right) e^{-jk_{x,0}s} \left(e^{-jk_{2z,m}(z-\frac{g}{2})} + e^{-j2h_2 k_{2z,m}} e^{+jk_{2z,m}(z-\frac{g}{2})}\right). \quad (A11)$$

where the subscripts 1 and 2 stand for the lower and upper regions, respectively. The parameters $C_{1m}$ and $C_{2m}$ represent unknown amplitude coefficients, $m$ is an integer, $k_1$ and $k_2$ are the corresponding fluid wave numbers, while $k_{1z,m} = \sqrt{k_1^2 - (m\pi/a)^2}$ and $k_{2z,m} = \sqrt{k_2^2 - (m\pi/a)^2}$. Normalized modal functions are given as

$$\sqrt{\frac{2-\delta_{m0}}{a}} \cos\left(\frac{m\pi}{a}x\right). \tag{A12}$$

where $\delta_{m0}$ is equal to zero unless $m = 0$.

One should note that the part of the Eqs. (A6)-(A11) given in brackets are the consequence of the reflection at the corrugation walls at $z = h_1 + \frac{g}{2}$ and $z = -h_2 - \frac{g}{2}$, where the reflection coefficients can be expressed as

$$\Gamma\left(\frac{g}{2} + h_2\right) = +1e^{-j2h_2 k_{z,m}}, \tag{A13}$$

$$\Gamma\left(-\frac{g}{2} - h_2\right) = +1e^{-j2h_1 k_{z,m}}. \tag{A14}$$

Boundary conditions for acoustic fields require continuity of pressure and normal velocity component at the boundaries $z = \pm\frac{g}{2}$ and zero normal velocity field on the rigid surface, which can be formulated as

$$\begin{aligned}
&At \begin{cases} z = \frac{g}{2} \\ s < x < s + a \end{cases} \begin{cases} p_2 = p_{gap} \\ v_{2z} = v_{gapz} \end{cases}, \\
&At \begin{cases} z = \frac{g}{2} \\ s + a < x < s + d \end{cases} \{v_{gapz} = 0, \\
&At \begin{cases} z = -\frac{g}{2} \\ 0 < x < a \end{cases} \begin{cases} p_1 = p_{gap} \\ v_{1z} = v_{gapz} \end{cases}, \\
&At \begin{cases} z = -\frac{g}{2} \\ a < x < d \end{cases} \{v_{gapz} = 0,
\end{aligned} \tag{A15}$$

Thus, the boundary conditions imply

$$\sum_{m=0}^{\infty} \frac{\eta_2 k_2}{k_{2z,m}} C_{2m} \sqrt{\frac{2-\delta_{m0}}{a}} \cos\left(\frac{m\pi}{a}(x-s)\right) e^{-jk_{x,0}s}\left(1 + e^{-j2h_2 k_{2z,m}}\right) =$$
$$\frac{1}{d}\sum_{p=-\infty}^{\infty} \frac{j\eta_{gap} k_{gap}}{k_{z,p}}\left(-B_p \sin(k_{z,p}\frac{g}{2}) + A_p \cos(k_{z,p}\frac{g}{2})\right) e^{-jk_{x,p}x}, \tag{A16}$$

$$\sum_{m=0}^{\infty} -\frac{\eta_1 k_1}{k_{1z,m}} C_{1m} \sqrt{\frac{2-\delta_{m0}}{a}} \cos\left(\frac{m\pi}{a}x\right)\left(1 + e^{-j2h_1 k_{1z,m}}\right) =$$
$$\frac{1}{d}\sum_{p=-\infty}^{\infty} \frac{j\eta_{gap} k_{gap}}{k_{z,p}}\left(B_p \sin(k_{z,p}\frac{g}{2}) + A_p \cos(k_{z,p}\frac{g}{2})\right) e^{-jk_{x,p}x}, \tag{A17}$$

$$\sum_{m=0}^{\infty} C_{2m} \sqrt{\frac{2-\delta_{m0}}{a}} \cos\left(\frac{m\pi}{a}(x-s)\right) e^{-jk_{x,0}s}\left(1 - e^{-j2h_2 k_{2z,m}}\right) = \quad \text{(A18)}$$

$$\frac{1}{d}\sum_{p=-\infty}^{\infty}\left(A_p \sin(k_{z,p}\tfrac{g}{2}) + B_p \cos(k_{z,p}\tfrac{g}{2})\right)e^{-jk_{x,p}x},$$

$$\sum_{m=0}^{\infty} C_{1m}\sqrt{\frac{2-\delta_{m0}}{a}} \cos\left(\frac{m\pi}{a}x\right)\left(1 - e^{-j2h_1 k_{1z,m}}\right) = \frac{1}{d}\sum_{p=-\infty}^{\infty}\left(-A_p \sin(k_{z,p}\tfrac{g}{2}) + \quad \text{(A19)}$$

$$B_p \cos(k_{z,p}\tfrac{g}{2})\right)e^{-jk_{x,p}x}.$$

Projecting Eqs. (A18) and (A19) on $e^{-jk_{x,p}x}$ over $s < x < d + s$ and $0 < x < d$, respectively, the following equations are obtained

$$\sum_{m=0}^{\infty} C_{1m}\left(1 - e^{-j2h_1 k_{1z,m}}\right) \widehat{\phi}_m(k_{x,p}) = \left(-A_p \sin(k_{z,p}\tfrac{g}{2}) + B_p \cos(k_{z,p}\tfrac{g}{2})\right), \quad \text{(A20)}$$

$$\sum_{m=0}^{\infty} C_{2m}\left(1 - e^{-j2h_2 k_{2z,m}}\right)\widehat{\phi}_m(k_{x,p}) e^{+j\frac{2\pi p}{d}s} = \left(A_p \sin(k_{z,p}\tfrac{g}{2}) + B_p \cos(k_{z,p}\tfrac{g}{2})\right), \quad \text{(A21)}$$

where

$$\int \sqrt{\frac{2-\delta_{m0}}{a}} \cos\left(\frac{m\pi}{a}x\right) e^{+jk_{x,p}x} = \widehat{\phi}_m(k_{x,p}), \quad \text{(A22)}$$

Consequently, the unknown amplitude coefficients can be expressed as

$$A_p = \frac{1}{2}\frac{-1}{\sin(k_{z,p}\tfrac{g}{2})}\left(\sum_m \widehat{\phi}_m(k_{x,p})\left(C_{1m}\left(1 - e^{-j2h_1 k_{1z,m}}\right) - C_{2m}\left(1 - \right.\right.\right.$$

$$\left.\left.\left. e^{-j2h_2 k_{2z,m}}\right)e^{+j\frac{2\pi p}{d}s}\right)\right), \quad \text{(A23)}$$

$$B_p = \frac{1}{2}\frac{1}{\cos(k_{z,p}\tfrac{g}{2})}\left(\sum_m \widehat{\phi}_m(k_{x,p})\left(C_{1m}\left(1 - e^{-j2h_1 k_{1z,m}}\right) + C_{2m}\left(1 - \right.\right.\right.$$

$$\left.\left.\left. e^{-j2h_2 k_{2z,m}}\right)e^{+j\frac{2\pi p}{d}s}\right)\right). \quad \text{(A24)}$$

Projecting Eqs. (A16) and (A17) on modal functions over $s < x < d + s$ and $0 < x < d$ respectively, the following equations are obtained

$$-\frac{\eta_1 k_1}{k_{1z,m}} \delta_{mm'} C_{1m}\left(1 + e^{-j2h_1 k_{1z,m}}\right)$$

$$= \frac{1}{d} \sum_{p=-\infty}^{\infty} \frac{j\eta_{gap} k_{gap}}{k_{z,p}} \left(B_p \sin\left(k_{z,p}\frac{g}{2}\right)\right. \tag{A25}$$

$$\left. + A_p \cos\left(k_{z,p}\frac{g}{2}\right)\right) \widehat{\phi_m}(-k_{x,p}),$$

$$\frac{\eta_2 k_2}{k_{2z,m}} \delta_{mm'} C_{2m}\left(1 + e^{-j2h_2 k_{2z,m}}\right) = \frac{1}{d}\sum_{p=-\infty}^{\infty}\frac{j\eta_{gap}k_{gap}}{k_{z,p}}\left(-B_p \sin(k_{z,p}\tfrac{g}{2}) + \right.$$

$$\left. A_p \cos(k_{z,p}\tfrac{g}{2})\right)\widehat{\phi_m}(-k_{x,p})e^{-j\frac{2\pi p}{d}s}. \tag{A26}$$

where $\delta_{mm'}$ is equal to zero unless $m = m'$.

When Eqs. (A23) and (A24) are combined with Eqs. (A25) and (A26), the following is obtained

$$\frac{jd}{k_{1z,m}}\frac{\eta_1 k_1}{\eta_{gap}k_{gap}} \delta_{mm'} C_{1m}\left(1 + e^{-j2h_1 k_{1z,m}}\right) =$$

$$\sum_p \sum_m \frac{\widehat{\phi_m}(k_{x,p})\widehat{\phi_m}(-k_{x,p})}{k_{z,p}}\left[\frac{1}{2}\tan\left(k_{z,p}\frac{g}{2}\right)\left(C_{1m}(1-e^{-j2h_1 k_{1z,m}})+C_{2m}(1-\right.\right.$$

$$\left.e^{-j2h_2 k_{2z,m}})e^{+j\frac{2\pi p}{d}s}\right) - \frac{1}{2}\cot\left(k_{z,p}\frac{g}{2}\right)\left(C_{1m}(1-e^{-j2h_1 k_{1z,m}}) - C_{2m}(1 - \right. \tag{A27}$$

$$\left.\left.e^{-j2h_2 k_{2z,m}})e^{+j\frac{2\pi p}{d}s}\right)\right],$$

$$\frac{-jd}{k_{2z,m}}\frac{\eta_2 k_2}{\eta_{gap}k_{gap}} \delta_{mm'} C_{2m}\left(1 + e^{-j2h_2 k_{2z,m}}\right) =$$

$$\sum_p \sum_m \frac{\widehat{\phi_m}(k_{x,p})\widehat{\phi_m}(-k_{x,p})}{k_{z,p}}e^{-j\frac{2\pi p}{d}s}\left[-\frac{1}{2}\tan\left(k_{z,p}\frac{g}{2}\right)\left(C_{1m}(1-\right.\right.$$

$$\left.e^{-j2h_1 k_{1z,m}})+C_{2m}(1-e^{-j2h_2 k_{2z,m}})e^{+j\frac{2\pi p}{d}s}\right) - \frac{1}{2}\cot\left(k_{z,p}\frac{g}{2}\right)\left(C_{1m}(1-\right. \tag{A28}$$

$$\left.\left.e^{-j2h_1 k_{1z,m}}) - C_{2m}(1-e^{-j2h_2 k_{2z,m}})e^{+j\frac{2\pi p}{d}s}\right)\right].$$

If the analysis is simplified so that lower and upper surfaces are considered the same, i.e. $h_1 = h_2 = h$, $\eta_1 = \eta_2 = \eta$, $k_1 = k_2 = k$, and $k_{1z,m} = k_{2z,m} = k_{z,m} = \sqrt{k^2 - (m\pi/a)^2}$ then Eqs. (A27) and (A28) can be reduced to

$$\frac{-jd}{k_{z,m}}\frac{\eta k}{\eta_{gap}k_{gap}}\delta_{mm'}C_{1m}(1+e^{-j2hk_{z,m}}) = \sum_p \sum_m \frac{1}{2}\frac{\widehat{\phi_m}(k_{x,p})\widehat{\phi_m}(-k_{x,p})}{k_{z,p}}(1-$$

$$e^{-j2hk_{z,m}})\left[-\tan\left(k_{z,p}\frac{g}{2}\right)\left(C_{1m}+C_{2m}e^{+j\frac{2\pi p}{d}s}\right)+\cot\left(k_{z,p}\frac{g}{2}\right)\left(C_{1m}-C_{2m}e^{+j\frac{2\pi p}{d}s}\right)\right], \quad (A29)$$

$$\frac{jd}{k_{z,m}}\frac{\eta k}{\eta_{gap}k_{gap}}\delta_{mm'}C_{2m}(1+e^{-j2hk_{z,m}}) = \sum_p \sum_m \frac{1}{2}\frac{\widehat{\phi_m}(k_{x,p})\widehat{\phi_m}(-k_{x,p})}{k_{z,p}}e^{-j\frac{2\pi p}{d}s}(1-$$

$$e^{-j2hk_{z,m}})\left[\tan\left(k_{z,p}\frac{g}{2}\right)\left(C_{1m}+C_{2m}e^{+j\frac{2\pi p}{d}s}\right)+\cot\left(k_{z,p}\frac{g}{2}\right)\left(C_{1m}-C_{2m}e^{+j\frac{2\pi p}{d}s}\right)\right]. \quad (A30)$$

The previous equations can be reformulated in a matrix form

$$\begin{bmatrix} M_{11} & M_{12} \\ M_{21} & M_{22} \end{bmatrix} \begin{bmatrix} C_{1m} \\ C_{2m} \end{bmatrix} = 0, \quad (A31)$$

and by imposing the matrix determinant to be equal to zero, solutions for the dispersion diagram can be obtained.